\documentclass[twocolumn,english,aps,prl,amsmath,amssymb,nofootinbib,showpacs,superscriptaddress]{revtex4-1}
\usepackage[latin9]{inputenc}
\usepackage{textcomp}
\usepackage{mathrsfs}
\usepackage{amsmath}
\usepackage{amssymb}
\usepackage{graphicx}
\usepackage{esint}

\makeatletter
\usepackage[english]{babel}
\usepackage{latexsym}
\usepackage{graphics}
\usepackage{epsfig}
\usepackage{color}
\usepackage{bm}

\usepackage{babel}

\usepackage{babel}

\makeatother

\usepackage{babel}
\begin{document}

\title{Superfluid density and critical velocity near the fermionic Berezinskii-Kosterlitz-Thouless
transition}

\author{Brendan C. Mulkerin}

\affiliation{Centre for Quantum and Optical Science, Swinburne University of Technology,
Melbourne 3122, Australia}

\author{Lianyi He}

\affiliation{State Key Laboratory of Low-Dimensional Quantum Physics and Department
of Physics, Tsinghua University, Beijing 100084, China}

\author{Paul Dyke}

\affiliation{Centre for Quantum and Optical Science, Swinburne University of Technology,
Melbourne 3122, Australia}

\author{Chris J. Vale}

\affiliation{Centre for Quantum and Optical Science, Swinburne University of Technology,
Melbourne 3122, Australia}

\author{Xia-Ji Liu}

\affiliation{Centre for Quantum and Optical Science, Swinburne University of Technology,
Melbourne 3122, Australia}

\affiliation{Kavli Institute for Theoretical Physics, UC Santa Barbara, USA}

\author{Hui Hu}

\affiliation{Centre for Quantum and Optical Science, Swinburne University of Technology,
Melbourne 3122, Australia}

\date{\today}
\begin{abstract}
We theoretically investigate superfluidity in a strongly interacting
Fermi gas confined to two dimensions at finite temperature. Using
a Gaussian pair fluctuation theory in the superfluid phase, we calculate
the superfluid density and determine the critical temperature and
chemical potential at the Berezinskii-Kosterlitz-Thouless transition.
We propose that the transition can be unambiguously demonstrated in
cold-atom experiments by stirring the superfluid Fermi gas using a
red detuned laser beam, to identify the characteristic jump in the
local Landau critical velocity at the superfluid-normal interface,
as the laser beam moves across the cloud. 
\end{abstract}


\maketitle
In two-dimensional (2D) many-body systems, topologically nontrivial
vortex fluctuations, that are suppressed due to vortex/anti-vortex
binding at low temperature, become amplified above a certain critical
temperature, leading to the so-called Berezinskii-Kosterlitz-Thouless
(BKT) transition \cite{Berezinskii1972,KT1973,Nelson1977}. The BKT
transition has been of great importance in different branches of physics
and has been observed in a range of settings \cite{Bishop1978,Resnick1981,Safonov1998,Hadzibabic2006}.
In particular, ultracold atomic gases are an ideal candidate to understand
the interaction-driven BKT physics \cite{Hadzibabic2006}, owing to
the unprecedented controllability over interatomic interactions, dimensionality
and species \cite{Bloch2008}. Over the past decade, the BKT transition
in a 2D weakly interacting Bose gas has been extensively studied by
measuring the phase coherence \cite{Hadzibabic2006,Clade2009}, confirming
the universal equation of state \cite{Hung2011,Yefsah2011}, probing
the superfluidity \cite{Desbuquois2012}, or observing the free vortex
proliferation \cite{Hadzibabic2006,Schweikhard2007,Choi2013}.

A 2D interacting Fermi gas at the crossover from a Bose-Einstein condensate
(BEC) to a Bardeen-Cooper-Schrieffer (BCS) superfluid provides a unique
platform to address the \emph{universal} BKT mechanism \cite{Zhang2008,Salasnich2013},
since the underlying character of the system changes from tightly
bound composite bosons to loosely bound Cooper pairs of fermions,
with decreasing attractions \cite{SchmittRink1989}. Indeed, the fermionic
BKT transition is now being pursued by several cold-atom laboratories
\cite{Martiyanov2010,Feld2011,Frohlich2011,Dyke2011,Orel2011,Koschorreck2012,Sommer2012,Zhang2012,Makhalov2014,Ong2015,Ries2015,Murthy2015,Dyke2016,Martiyanov2016,Fenech2016,Boettcher2016,Cheng2016,Turlapov2017},
and there are indications of the transition from the measurements
of pair condensation and correlation function, where: (i) the center-of-mass
momentum distribution of Cooper pairs, $n_{\mathbf{Q}}$, exhibits
anomalous enhancement near $\mathbf{Q}=0$ below a certain temperature
\cite{Ries2015}, and (ii) the first-order correlation function $g_{1}(r)$
in real space decays algebraically \cite{Murthy2015}. However, confirmation
of the transition is still to be demonstrated, as these two features
may be explained using a strong-coupling theory in the normal phase
\cite{Matsumoto2016}. This situation marks the importance of having
accurate theoretical predictions for the fermionic BKT transition.

The purpose of this Letter is to apply a strong-coupling theory, beyond
mean-field, to a 2D interacting Fermi gas in the \emph{superfluid}
phase and present \emph{semi-quantitative} predictions for the BKT
critical chemical potential, critical temperature and the critical
velocity at the whole BEC-BCS crossover. Through a fully \emph{microscopic}
calculation of both superfluid density and critical velocity, beyond
the phenomenological Landau quasi-particle picture, we predict the
occurrence of a significant discontinuity in the critical velocity
across the transition as a result of the universal jump in superfluid
density \cite{KT1973}, which would provide an unambiguous proof of
the fermionic BKT transition.

\begin{figure}
\centering{}\includegraphics[width=0.45\textwidth]{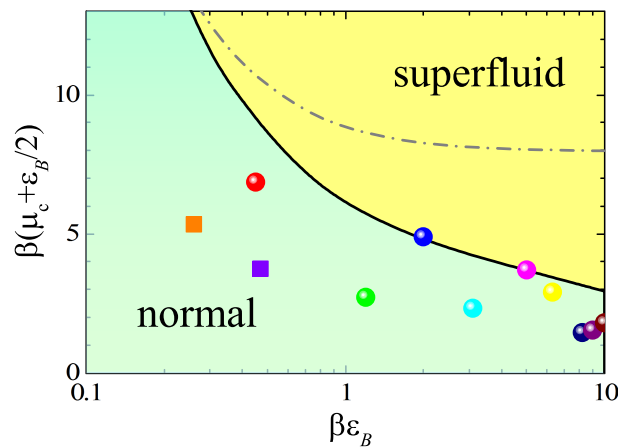} \caption{(color online). The critical chemical potential (with $\varepsilon_{B}/2$
added) as a function of the interaction strength. The black solid
line and the grey dot-dashed line show the GPF and mean-field results,
respectively. The symbols (in different colors) show the largest chemical
potential achieved in the recent density equation of state measurements
\cite{Fenech2016,Boettcher2016}, at different interaction strengths.
\label{fig:criticalmu} }
\end{figure}

\begin{figure}
\centering{}\includegraphics[width=0.45\textwidth]{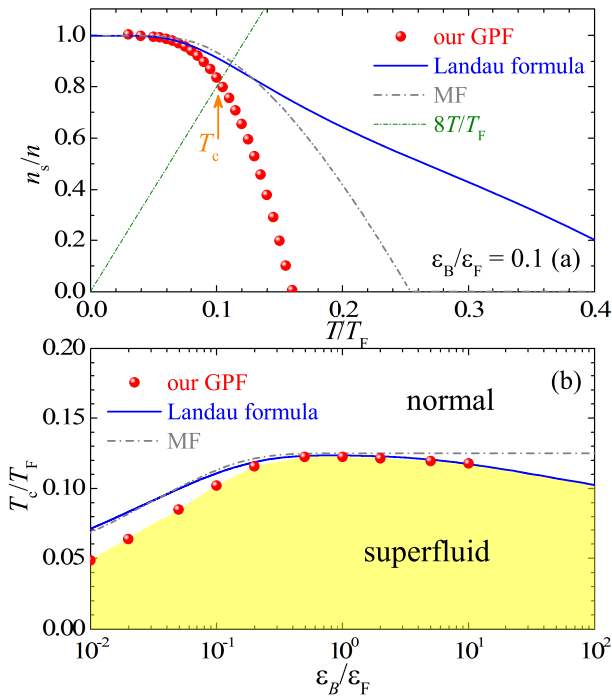} \caption{(color online). (a) The superfluid fraction as a function of temperature
at interaction strength $\varepsilon_{B}=0.1\varepsilon_{F}$. Our
GPF prediction (red circles) is compared with the mean-field result
(grey dot-dashed line) and the approximated result based on the zero-temperature
GPF (blue solid line) \cite{Bighin2016}. The intersection with the
curve $8T/T_{F}$ determines the BKT transition temperature. (b) The
critical temperature as a function of $\varepsilon_{B}/\varepsilon_{F}$.
\label{fig:Tc}}
\end{figure}

The theoretical description of pairing in a 2D interacting Fermi gas
at finite temperature is a long-standing challenge due to strongly
enhanced quantum and thermal fluctuations. There have been intense
theoretical efforts over the last thirty years, to understand the
corresponding mechanism in 2D layered high-temperature superconductors
\cite{SchmittRink1989,Loktev2001,Lee2006}. To a large extent, current
knowledge of the fermionic BKT transition builds on mean-field approach
\cite{Zhang2008,Salasnich2013}, which breaks down when interactions
become stronger. There are a number of studies that take into account
strong pair fluctuations based on the many-body $T$-matrix scheme
\cite{Klimin2012,Watanabe2013,Bauer2014,Marsiglio2015,He2015,Mulkerin2015,Bighin2016},
however, these calculations typically focus on the normal state due
to technical difficulties. The ab-initio quantum Monte Carlo (QMC)
simulations at finite temperature encounter similar issues \cite{Anderson2015}.
In this Letter, we consider a Gaussian pair fluctuation (GPF) theory
\cite{Hu2006,Hu2007,Diener2008}, which is known to provide a reliable
2D equation of state at zero temperature \cite{He2015}. We generalize
the GPF theory for finite temperatures below the superfluid transition,
solving a crucial technical problem of removing divergences in numerics.
This enables us to calculate the superfluid density, the key quantity
in characterizing the BKT transition, beyond the mean-field and taking
into account quantum fluctuations. Our main results, as shown in Fig.
\ref{fig:criticalmu} and Fig. \ref{fig:Tc}(b), are of significant
importance for further BKT experiments with cold fermions.

\textit{The GPF theory at finite T}. \textemdash{} A 2D interacting
Fermi gas is well-described by the Hamiltonian \cite{SchmittRink1989},
\begin{equation}
\mathcal{H}=\sum_{\sigma}\bar{\psi}_{\sigma}(\mathbf{r})\mathcal{H}_{0}\psi_{\sigma}(\mathbf{r})-U\bar{\psi}_{\uparrow}(\mathbf{r})\bar{\psi}_{\downarrow}(\mathbf{r})\psi_{\downarrow}(\mathbf{r})\psi_{\uparrow}(\mathbf{r}),\label{eq:scHami}
\end{equation}
where $\psi_{\sigma}(\mathbf{r})$ is the annihilation operator for
the spin state $\sigma=\uparrow,\downarrow$, $\mathcal{H}_{0}=-\hbar^{2}\nabla^{2}/(2M)-\mu$
the kinetic Hamiltonian with atomic mass $M$, $\mu$ the chemical
potential, and $U$ denotes the bare interaction strength of a contact
interaction between unlike fermions and is related to the binding
energy $\varepsilon_{B}$ via, $1/U=\sum_{\mathbf{k}}(\hbar^{2}\mathbf{k}^{2}/M+\varepsilon_{B})^{-1}$.

Technical details of the GPF theory have been extensively discussed
elsewhere \cite{He2015,Hu2006,Diener2008}, here, we only present
a brief overview of the key equations and refer the readers to Supplementary
Material for further details \cite{supp}. Within the GPF framework,
we account for strong pair fluctuations at the Gaussian level, beyond
the standard mean-field treatment, and consider separately their contributions
to the thermodynamic potential, $\Omega=\Omega_{\textrm{MF}}+\Omega_{\textrm{GF}}$.
These two parts can be represented by the BCS Green's function $\mathscr{\mathcal{G}}_{0}(\mathbf{k},i\omega_{m})$
and the vertex function $\Gamma(\mathbf{q},$$i\nu_{l})$ (i.e., the
Green's function of Cooper pairs): $\Omega_{\textrm{MF}}=-k_{B}T\sum_{\mathbf{k},i\omega_{m}}\ln[-\mathcal{G}_{0}^{-1}]$,
and $\Omega_{\textrm{GF}}=(k_{B}T/2)\sum_{\mathbf{q},i\nu_{l}}\ln[-\Gamma^{-1}]$.
That is, the expressions of the thermodynamic potentials for ideal
fermions and bosons, where $\omega_{m}=(2m+1)\pi k_{B}T$ and $\nu_{l}=2\pi lk_{B}T$
are the fermionic and bosonic Matsubara frequencies with integers
$m$ and $l$, respectively. In other words, the system may be viewed
as a \emph{non-interacting} mixture of fermions and pairs. Though
the picture is simple, it captures the essential physics for weak
and strong interactions. Indeed, at zero temperature, the GPF theory
provides a quantitative description of the BEC-BCS crossover in both
3D \cite{Hu2006,Hu2007,Diener2008} and 2D \cite{He2015,Shi2015}.
This can be extended straight forwardly to the general situation where
the condensed pairs flow with a wavevector $\mathbf{Q}$, as represented
by a pairing gap $\Delta e^{i\mathbf{Q}\cdot\mathbf{r}}$ \cite{Taylor2006,supp}.
In this case, 
\begin{equation}
\Omega_{\textrm{MF}}=\frac{\Delta^{2}}{U}+\sum_{\mathbf{k}}\left[\tilde{\xi}_{\mathbf{k}}-E_{\mathbf{k}}-\frac{2}{\beta}\ln\left(1+e^{\beta E_{\mathbf{k}}^{+}}\right)\right],\label{OmegaMF}
\end{equation}
where $\smash{\tilde{\xi}_{\mathbf{k}}\equiv\hbar^{2}\mathbf{k}^{2}/(2M)-[\mu-\hbar^{2}\mathbf{Q}^{2}/(8M)]}$,
$E_{\mathbf{k}}\equiv\sqrt{\tilde{\xi}_{\mathbf{k}}^{2}+\Delta^{2}}$,
$\beta=1/(k_{B}T)$ and $E_{\mathbf{k}}^{\pm}\equiv E_{\mathbf{k}}\pm\hbar^{2}\mathbf{k}\cdot\mathbf{Q}/(2M)$.
The expression for the thermodynamic potential of pair fluctuations
is more subtle \cite{Hu2006,Diener2008}, 
\begin{eqnarray}
\Omega_{\textrm{GF}} & = & k_{B}T\sum_{\mathcal{Q}\equiv\left(\mathbf{q},i\nu_{l}\right)}\mathcal{S}\left(\mathcal{Q}\right)e^{i\nu_{l}0^{+}},\label{OmegaGF}\\
\mathcal{S}\left(\mathcal{Q}\right) & = & \frac{1}{2}\ln\left[1-\frac{M_{12}^{2}\left(\mathcal{Q}\right)}{M_{11}\left(\mathcal{Q}\right)M_{11}\left(-\mathcal{Q}\right)}\right]+\ln M_{11}\left(\mathcal{Q}\right),\nonumber 
\end{eqnarray}
and the matrix elements $M_{11}\left(\mathcal{Q}\right)$ and $M_{12}\left(\mathcal{Q}\right)$
are given in \cite{supp}. The density $n$ of the system can be calculated
using $n=-\partial(\Omega_{\textrm{MF}}+\Omega_{\textrm{GF}})/\partial\mu$,
which determines the Fermi wavevector $k_{F}=(2\pi n)^{1/2}$, energy
$\varepsilon_{F}=\pi n\hbar^{2}/M$ and temperature $T_{F}=\varepsilon_{F}/k_{B}$. 

\begin{figure}
\begin{centering}
\includegraphics[width=0.45\textwidth]{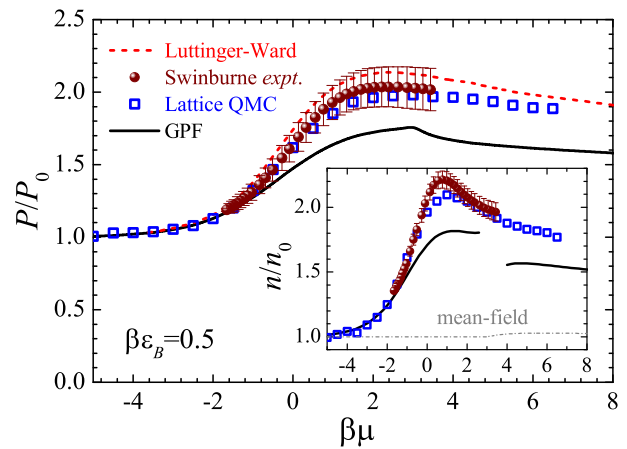} 
\par\end{centering}
\caption{(color online). Pressure equation of state at $\beta\varepsilon_{B}=0.5$.
The prediction from the GPF theory (black solid line) is compared
with the results from the Luttinger-Ward theory (red dashed line)
\cite{Bauer2014,Mulkerin2015} and lattice QMC simulation (blue squares)
\cite{Anderson2015}, and the experimental data from Swinburne \cite{Fenech2016}
(solid circles with error bar at a slightly smaller $\beta\varepsilon_{B}=0.47$).
The inset shows the density equation of state at the same interaction
strength. Here, $P_{0}(\mu)$ and $n_{0}(\mu)$ are the pressure and
density of an ideal Fermi gas, respectively. \label{fig:EoS}}
\end{figure}

Despite the simplicity and elegance of the GPF theory, it is not easy
to solve numerically in general. The technical difficulty comes from
the sum over the bosonic Matsubara frequency $i\nu_{l}$ in Eq.~(\ref{OmegaGF}),
which is divergent. For an interacting 2D Fermi gas at zero temperature
the problem may be solved by utilizing an additional function which
has no singularities or zeros in the left hand-plane \cite{He2015,Diener2008}.
At finite temperature, however, the GPF has only been approximately
treated by taking into account the effects of low-energy phonon modes
\cite{Klimin2012,Bighin2016}. Here, we overcome the divergence by
writing \cite{Tempere2008}, 
\begin{equation}
\frac{1}{\beta}\sum_{\left|l\right|>l_{0}}\mathcal{S}_{\eta}\left(\mathbf{q},i\nu_{l}\right)=-\frac{1}{\pi}\int_{-\infty}^{+\infty}d\omega\frac{\textrm{Im}\mathcal{S_{\eta}}\left(\mathbf{q},\omega+i\gamma\right)}{e^{\beta\omega}+1},\label{TempereTrick}
\end{equation}
where $\mathcal{S}_{\eta}(\mathbf{q},i\nu_{l})\equiv\mathcal{S}(\mathbf{q},i\nu_{l})e^{i\nu_{l}\eta}$
and $\gamma=(2l_{0}+1)\pi/\beta$ for arbitrary positive integer $l_{0}$.
Thus, the contribution to $\Omega_{\textrm{GF}}$ at a given $\mathbf{q}$
can be calculated by using Eq. (\ref{TempereTrick}) and taking the
remaining discrete sum with $\left|l\right|<l_{0}$, in the limit
of $\eta\rightarrow0^{+}$. We have confirmed that this numerical
procedure is robust and independent of the choice of $l_{0}$.

To illustrate the importance of our \emph{full} treatment of the GPF,
we show in Fig.~\ref{fig:EoS} the results for the pressure and density
equations of state at interaction strength $\smash{\beta\varepsilon_{B}=0.5}$
with $\smash{\mathbf{Q}=0}$, compared with the predictions from the
mean-field theory, above $T_{c}$ calculations with the self-consistent
Luttinger-Ward theory \cite{Bauer2014,Mulkerin2015} and lattice QMC
simulation \cite{Anderson2015}, and with recent experimental measurements
\cite{Fenech2016}. It is reasonable from the comparison of results
in Fig.~\ref{fig:EoS} that the GPF theory is semi-quantitatively
reliable over the whole temperature regime. For a \emph{superfluid}
2D Fermi gas, the GPF theory provides the best description to date,
as current mean-field theories strongly under-estimate the interaction
effects \cite{Salasnich2013} and there are no superfluid QMC calculations
at finite temperature. Alternative $T$-matrix theories have so far
focused on the normal state only and predicted a 2D superfluid transition
at \emph{zero} temperature \cite{Marsiglio2015}.

\textit{Superfluid density and phase diagrams.} \textemdash{} We now
consider the case that the condensed pairs flow with superfluid velocity
$\mathbf{v}_{s}=\hbar\mathbf{Q}/(2M)$. Treating $\mathbf{v}_{s}$
as small, the superfluid density $n_{s}$ of the system can be calculated
from the lowest-order change in the thermodynamic potential, i.e.,
$\Delta\Omega=\Omega(\mathbf{v}_{s})-\Omega(0)\simeq Mn_{s}\mathbf{v}_{s}^{2}$/2,
due to the added kinetic energy of the superfluid flow \cite{Taylor2006},
thus, we obtain, 
\begin{equation}
n_{s}=\frac{1}{M}\left[\frac{\partial^{2}\Omega\left(\mathbf{v}_{s}\right)}{\partial v_{s}^{2}}\right]_{v_{s}=0}=\frac{4M}{\hbar^{2}}\left[\frac{\partial^{2}\Omega\left(\mathbf{Q}\right)}{\partial Q^{2}}\right]_{Q=0}.
\end{equation}
The BKT critical temperature $T_{c}$ can then be estimated by self-consistently
solving the KT criterion \cite{Nelson1977,Salasnich2013}, 
\begin{equation}
k_{B}T_{c}=\frac{\pi}{2}\frac{\hbar^{2}}{4M}n_{s}\left(T_{c}\right).\label{KT}
\end{equation}
\begin{figure}
\centering{}\includegraphics[width=0.45\textwidth]{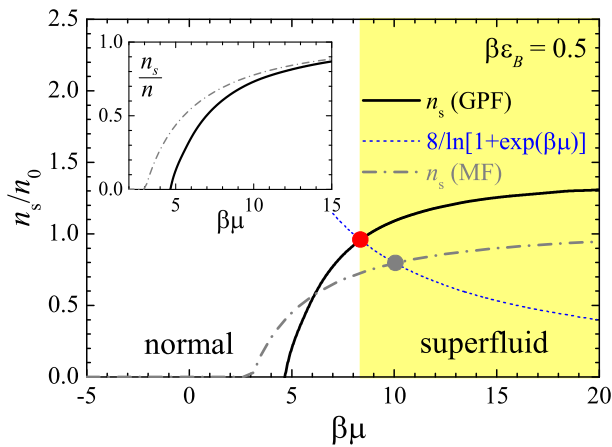} \caption{(color online). The superfluid density, in units of the density of
an ideal Fermi gas $n_{0}$, as a function of the chemical potential
at the interaction strength $\beta\varepsilon_{B}=0.5$. The GPF and
mean-field predictions are shown by the black solid and grey dot-dashed
lines, respectively. The circles indicate the critical superfluid
density (or chemical potential) for the BKT transition. The inset
shows the superfluid fraction $n_{s}/n$. \label{fig:sf}}
\end{figure}

Figure~\ref{fig:sf} reports the superfluid density $n_{s}$ at the
interaction strength $\beta\varepsilon_{B}=0.5$, as a function the
dimensionless chemical potential $\beta\mu$. The main figure shows
$n_{s}$ in units of the density of an ideal Fermi gas $n_{0}=2\lambda_{T}^{-2}\ln(1+e^{\beta\mu})$,
where $\lambda_{T}\equiv\sqrt{2\pi\hbar^{2}/(Mk_{B}T)}$ is the thermal
wavelength, while the inset shows the superfluid fraction $n_{s}/n$.
For comparison, we also plot the mean-field results (dot-dashed).
By dividing both sides of the KT criterion, Eq. (\ref{KT}), by $n_{0}$,
we find that the dimensionless critical chemical potential, $(\beta\mu)_{c}$,
may be obtained by plotting $n_{s}/n_{0}$ and looking for the intercept
with $8/\ln(1+e^{\beta\mu})$. Towards the low-temperature regime,
$\beta\mu\rightarrow\infty$, the superfluid density calculated using
the mean-field theory is typically under-estimated, although the superfluid
fractions from both mean-field and GPF theories saturate to unity.
Consequently, the mean-field theory predicts a larger critical chemical
potential.

By repeating the calculations at different interaction strengths we
obtain a phase diagram for the critical chemical potential, as shown
in Fig. \ref{fig:criticalmu}. This phase diagram is particularly
useful for current cold-atom experiments, where the Fermi gas is confined
in a harmonic trapping potential, $V(\mathbf{r})$, and is inhomogeneous.
A section of the cloud is locally superfluid if its local chemical
potential $\mu_{\textrm{loc}}=\mu-V(\mathbf{r})$ is larger than $\mu_{c}$.
Therefore, experimentally, once the chemical potential at the trap
center, $\mu$, and the temperature, $T$, are measured by fitting
the density equation of state at the edge of the cloud with the known
virial expansion \cite{Fenech2016}, one can then determine the superfluid
radius of the Fermi cloud from our phase diagram, Fig.~\ref{fig:criticalmu}.
To make a close connection with experiments, in the figure we show
the largest chemical potential achieved in recent equation of state
measurements \cite{Fenech2016,Boettcher2016}. It is encouraging to
see that the experiment was approaching the BKT transition.

On the theoretical side, it is of interest to determine the phase
diagram for the parameter space of $T_{c}/T_{F}$ and $\varepsilon_{B}/\varepsilon_{F}$,
where, we calculate the superfluid fraction as a function of $T/T_{F}$.
A typical prediction at $\varepsilon_{B}/\varepsilon_{F}=0.1$ is
illustrated in Fig. \ref{fig:Tc}(a) by solid circles, contrasted
with the mean-field result (dot-dashed line). The superfluid density
of a 2D interacting Fermi gas has been recently calculated by Bighin
and Salasnich \cite{Bighin2016} using Landau's phenomenological formulation
for the normal density and the quasiparticle spectrum based on the
zero-temperature GPF equation of state \cite{Baym2013}. Their result
is plotted in Fig. \ref{fig:Tc}(a) for comparison. We find that the
prediction of Landau's formulation agrees well with our full GPF calculation
at low temperatures, where $n_{s}/n\sim1$, but significantly over-estimates
the superfluid fraction when the temperature becomes larger. According
to the KT criterion, the critical temperature $T_{c}/T_{F}$ can be
extracted by locating the intercept point between the curves $n_{s}/n$
and $8T/T_{F}$. The resulting phase diagram is reported in Fig. \ref{fig:Tc}(b).
Our result shows a significant improvement on the BCS side over the
previous theoretical predictions \cite{Salasnich2013,Bighin2016}.
While on the BEC side (i.e., $\varepsilon_{B}>0.5\varepsilon_{F}$),
our result follows closely to the approximate prediction from Landau's
formula, since in the latter, the superfluid fraction at low temperatures
$T\sim0.1T_{F}$ is reasonably approximated. In the deep BEC regime
our GPF result approaches the anticipated BKT critical temperature
of a weakly interacting Bose gas \cite{Bighin2016,Petrov2003}, since,
the molecular scattering length is correctly reproduced in the GPF
theory \cite{He2015,Salasnich2015}. In this respect, the phase diagram
Fig. \ref{fig:Tc}(b) gives a coherent picture across the whole BEC-BCS
crossover.

\textit{Probing the fermionic BKT transition.} \textemdash{} We now
consider way to unambiguously identify the fermionic BKT transition.
Due to strong interactions, measurements of both phase coherence and
free vortex proliferation, which are efficient for a weakly interacting
2D Bose gas, do not work well. Instead, we follow the idea of the
recent superfluidity measurement \cite{Desbuquois2012} and propose
to observe the superfluid behavior of an interacting 2D Fermi gas
by stirring the cloud with a red detuned laser beam. When the Fermi
cloud is in the superfluid state, we anticipate that the measured
critical velocity will have a sudden jump as the position of the stirred
beam moves across a critical radius $r_{c}$, which corresponds to
the critical chemical potential $\mu_{c}=\mu-V(r_{c})$. This sudden
increase is caused by the universal jump in the superfluid density,
since \emph{just} below (above) the BKT critical temperature (chemical
potential), the finite superfluid density is able to support nonzero
superfluid flow \cite{footnote}.

\begin{figure}
\centering{}\includegraphics[width=0.45\textwidth]{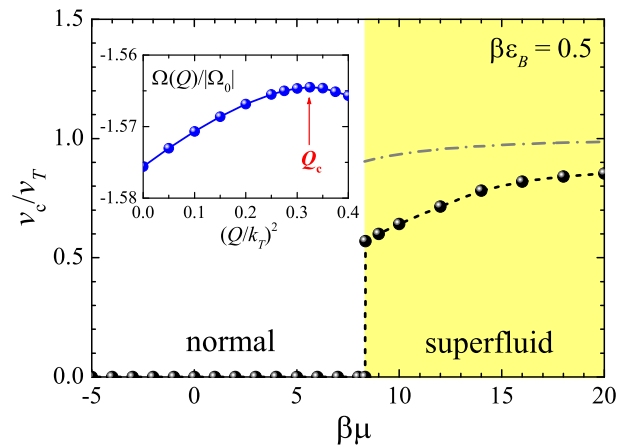} \caption{(color online). The critical velocity $v_{c}$, in units of the thermal
velocity $v_{T}\equiv(2k_{B}T/M)^{1/2}=\hbar k_{T}$, as a function
of $\beta\mu$ at the interaction strength $\beta\varepsilon_{B}=0.5$.
The black circles (with dashed line) and the grey dot-dashed line
show the GPF and mean-field predictions, respectively. The inset shows
the thermodynamic potential at nonzero superfluid velocity $v=\hbar Q/(2M)$,
which exhibits a local maximum at $v_{c}$. \label{fig:vc}}
\end{figure}

Theoretically, we calculate the critical velocity from the velocity
dependence of the thermodynamic potential $\Omega(\mathbf{v})$ at
a given temperature, $T$. With increasing superfluid flow, the loss
of stability of the system is indicated by the appearance of a local
maximum in the thermodynamic potential, as illustrated in the inset
of Fig. \ref{fig:vc}. The determined critical velocity at the interaction
strength $\beta\varepsilon_{B}=0.5$ is presented in the main figure.
The apparent discontinuity at $(\beta\mu)_{c}\sim8$ serves as a smoking-gun
signature for the BKT transition. To give some realistic numbers,
consider a single 2D cloud of $N=40,000$ neutral $^{6}$Li atoms
in a hybrid optical/magnetic trap with frequency $\omega_{x}\simeq\omega_{y}\sim2\pi\times25$
Hz at temperature $T\sim20$ nK and at binding energy $\varepsilon_{B}=10$
nK (satisfying $\beta\varepsilon_{B}\sim0.5$), which is within the
regime attainable at Swinburne \cite{Fenech2016}. The chemical potential
at the trap center is estimated to be $\mu\sim240$ nK. Thus, the
superfluid radius is about $r_{c}\sim100$ \textmu m, and from Fig.
\ref{fig:vc}, the anticipated jump in the critical velocity would
be about $\Delta v_{c}\simeq0.6v_{T}\sim4.5$ mm/s, which is readily
detectable \cite{Weimer2015}.

\textit{Conclusions}. \textemdash{} We have established reliable estimates
for the superfluid transition temperature of a strongly interacting
2D Fermi gas. This is done by developing a Gaussian pair fluctuation
theory that provides semi-quantitatively accurate predictions on the
superfluid density at any interaction strength and temperature. Our
results support on-going cold-atom experiments to unambiguously observe
the fermionic Berezinskii-Kosterlitz-Thouless transition. Our approach
may also be useful for understanding the superfluid phases of the
2D Hubbard model \cite{Paiva2004}. 
\begin{acknowledgments}
We are grateful to Joaquín Drut for sharing the QMC results in Ref.
\cite{Anderson2015} and to Yvan Castin and Gora Shlyapnikov for stimulating
discussions on the sudden jump in the critical velocity. This research
was supported under Australian Research Council's Discovery Projects
funding scheme (project numbers DP140100637 and DP140103231) and Future
Fellowships funding scheme (project numbers FT130100815 and FT140100003).
LH was supported by the Thousand Young Talents program in China. XJL
was supported in part by the National Science Foundation under Grant
No. NSF PHY-1125915, during her visit to KITP. 
\end{acknowledgments}

\section*{Supplemental Material: Details of the thermodynamic potential}

\noindent In greater detail we can derive the matrix elements of the
pair fluctuation thermodynamic potential, and for completeness we
also show the mean-field thermodynamic potential. Let us consider
the general situation where the condensed pairs flow with a wavevector
$\mathbf{Q}$, or superfluid velocity $\mathbf{v}_{s}=\hbar\mathbf{Q}/(2M)$,
as represented by a pairing gap $\Delta e^{i\mathbf{Q}\cdot\mathbf{r}}$
\cite{Taylor2006}. In this case, the mean-field thermodynamic potential
is given by \cite{Taylor2006}, 
\begin{equation}
\Omega_{\textrm{MF}}\left(\mathcal{Q}\right)=\frac{\Delta^{2}}{U}+\sum_{\mathbf{k}}\left[\tilde{\xi}_{\mathbf{k}}-E_{\mathbf{k}}-\frac{2}{\beta}\ln\left(1+e^{\beta E_{\mathbf{k}}^{+}}\right)\right],\label{OmegaMF_SM}
\end{equation}
where $\tilde{\xi}_{\mathbf{k}}\equiv\hbar^{2}\mathbf{k}^{2}/(2M)-[\mu-\hbar^{2}\mathbf{Q}^{2}/(8M)]$,
$E_{\mathbf{k}}\equiv\sqrt{\tilde{\xi}_{\mathbf{k}}^{2}+\Delta^{2}}$,
$\beta=1/(k_{B}T)$ and $E_{\mathbf{k}}^{\pm}\equiv E_{\mathbf{k}}\pm\hbar^{2}\mathbf{k}\cdot\mathbf{Q}/(2M)$,
and to ensure the gapless Goldstone mode, the pairing gap $\Delta$
should be calculated using the mean-field gap equation, 
\begin{equation}
\sum_{\mathbf{k}}\left[\frac{1-2f\left(E_{\mathbf{k}}^{+}\right)}{2E_{\mathbf{k}}}-\frac{1}{\hbar^{2}\mathbf{k}^{2}/M+\varepsilon_{B}}\right]=0,
\end{equation}
with the Fermi distribution function $f(x)\equiv1/(e^{\beta x}+1)$.
The expression for the thermodynamic potential of pair fluctuations
is more subtle \cite{Hu2006,Diener2008}: 
\begin{eqnarray}
\Omega_{\textrm{GF}}\left(\mathcal{Q}\right) & = & k_{B}T\sum_{\mathcal{Q}\equiv\left(\mathbf{q},i\nu_{l}\right)}\mathcal{S}\left(\mathcal{Q}\right)e^{i\nu_{l}0^{+}},\label{OmegaGF_SM}\\
\mathcal{S}\left(\mathcal{Q}\right) & = & \frac{1}{2}\ln\left[1-\frac{M_{12}^{2}\left(\mathcal{Q}\right)}{M_{11}\left(\mathcal{Q}\right)M_{11}\left(-\mathcal{Q}\right)}\right]+\ln M_{11}\left(\mathcal{Q}\right),\nonumber 
\end{eqnarray}
where the matrix elements of $-\Gamma^{-1}(\mathcal{Q})$ are given
by \cite{Taylor2006},\begin{widetext} 
\begin{eqnarray}
M_{11} & \left(\mathcal{Q}\right)= & \frac{1}{U}+\sum_{\mathbf{k}}\left[u_{+}^{2}u_{-}^{2}\frac{1-f_{+}^{\left(+\right)}-f_{-}^{\left(-\right)}}{i\tilde{\nu}_{l}-E_{+}-E_{-}}-u_{+}^{2}v_{-}^{2}\frac{f_{+}^{\left(+\right)}-f_{-}^{\left(+\right)}}{i\tilde{\nu}_{l}-E_{+}+E_{-}}+v_{+}^{2}u_{-}^{2}\frac{f_{+}^{\left(-\right)}-f_{-}^{\left(-\right)}}{i\tilde{\nu}_{l}+E_{+}-E_{-}}-v_{+}^{2}v_{-}^{2}\frac{1-f_{+}^{\left(-\right)}-f_{-}^{\left(+\right)}}{i\tilde{\nu}_{l}+E_{+}+E_{-}}\right],\nonumber \\
M_{12} & \left(\mathcal{Q}\right)= & \sum_{\mathbf{k}}\left(u_{+}v_{+}u_{-}v_{-}\right)\left[-\frac{1-f_{+}^{\left(+\right)}-f_{-}^{\left(-\right)}}{i\tilde{\nu}_{l}-E_{+}-E_{-}}-\frac{f_{+}^{\left(+\right)}-f_{-}^{\left(+\right)}}{i\tilde{\nu}_{l}-E_{+}+E_{-}}+\frac{f_{+}^{\left(-\right)}-f_{-}^{\left(-\right)}}{i\tilde{\nu}_{l}+E_{+}-E_{-}}+\frac{1-f_{+}^{\left(-\right)}-f_{-}^{\left(+\right)}}{i\tilde{\nu}_{l}+E_{+}+E_{-}}\right].\label{M11M12}
\end{eqnarray}
\end{widetext} Here, we use the short-hand notations $i\tilde{\nu_{l}}\equiv i\nu_{l}-\hbar^{2}\mathbf{q}\cdot\mathbf{Q}/(2M)$,
$E_{\pm}\equiv E_{\mathbf{k}\pm\mathbf{q}/2}$, $f_{\pm}^{\left(\pm\right)}\equiv f(E_{\mathbf{k}\pm\mathbf{q}/2}^{\pm})$,
$u_{\pm}^{2}=(1+\tilde{\xi}_{\mathbf{k}\pm\mathbf{q}/2}/E_{\mathbf{k}\pm\mathbf{q}/2})/2$
and $v_{\pm}^{2}=1-u_{\pm}^{2}$ .

\begin{thebibliography}{10}
\bibitem{footnote}This is particularly clear on the BEC side, where
the critical velocity is given by the sound velocity, according to
Landau's picture. The latter remains finite right at the BKT transition.
\end{thebibliography}
\end{document}